\documentclass[aps,nofootinbib,showpacs,preprintnumbers,twocolumn,superscriptaddress,prb]{revtex4-1}

\usepackage{amsmath,amssymb}
\usepackage{graphicx}
\usepackage{bm}
\usepackage{color}

\definecolor{michael}{rgb}{0.9,0.,0.1}

\newcommand{\I}{\mathrm{i}}
\newcommand{\E}{\mathrm{e}}

\begin{document}

\title{Quantum Monte Carlo studies of edge magnetism in chiral graphene nanoribbons}

\author{Michael Golor}
\affiliation{Institute for Theoretical Solid State Physics, RWTH Aachen University, Aachen, Germany}
\affiliation{JARA-HPC High Performance Computing}
\affiliation{JARA-FIT Fundamentals of Future Information Technology}
\author{Thomas C. Lang}
\affiliation{Department of Physics, Boston University, Boston, MA 02215, USA}
\affiliation{Institute for Theoretical Solid State Physics, RWTH Aachen University, Aachen, Germany}
\affiliation{JARA-HPC High Performance Computing}
\affiliation{JARA-FIT Fundamentals of Future Information Technology}
\author{Stefan Wessel}
\affiliation{Institute for Theoretical Solid State Physics, RWTH Aachen University, Aachen, Germany}
\affiliation{JARA-HPC High Performance Computing}
\affiliation{JARA-FIT Fundamentals of Future Information Technology}

\begin{abstract}
We investigate chiral graphene nanoribbons using projective quantum Monte Carlo simulations within the local Hubbard model description and study the effects of electron-electron interactions on the electronic and magnetic properties at the ribbons' edges. Static and dynamical properties are analyzed for nanoribbons of varying width and edge chirality, and compared to a self-consistent Hartee-Fock mean-field approximation. 
Our results show that for chiral ribbons of sufficient width, the spin correlations exhibit exceedingly long correlation lengths, even between zigzag segments that are well separated by periodic armchair regions. Characteristic enhancements in the magnetic correlations for distinct ribbon widths and chiralities are associated with energy gaps in the tight-binding limit of such ribbons. We identify specific signatures in the local density of states and low-energy modes in the local spectral function 
which  directly relate to 
enhanced electronic correlations along graphene nanoribbons. These signatures in the local density of states might be accessed by scanning tunneling spectroscopy on graphene nanoribbons. 
\end{abstract}

\pacs{71.10.Fd,71.27.+a,73.21.-b,73.22.Pr,75.70.Ak}

\maketitle

\section{Introduction}

Graphene nanoribbons (GNRs), laterally confined, nanometer wide one-dimensional long carbon strips, are currently intensively examined with respect to  their electronic properties and potential for future graphene-based electronic devices.\cite{Son06a,Chen07,Yazyev08,Schwierz10,Vazyev13} Various fabrication strategies have been explored recently in order to achieve high-quality GNRs, such as the unzipping of carbon nanotubes~\cite{Jiao09,Kosynkin09,Jiao10}, or the direct chemical synthesis of GNRs.\cite{Yang08,Cai10,Pan12,Huang12,Linden12} Transport measurements report sizable band gaps in GNRs,\cite{Li08} which in general depend on the edge geometry. In case of GNRs with armchair edges (aGNRs), these gaps are a consequence of the lateral quantum confinement, while for the highly symmetric GNRs with zigzag edges (zGNRs), an energy gap is theoretically predicted to arise due to the spontaneous spin polarization established along the zigzag edges.\cite{Fujita96,Son06a,Son06b}
\begin{figure}[t!]
\centering
  \includegraphics[width=\columnwidth]{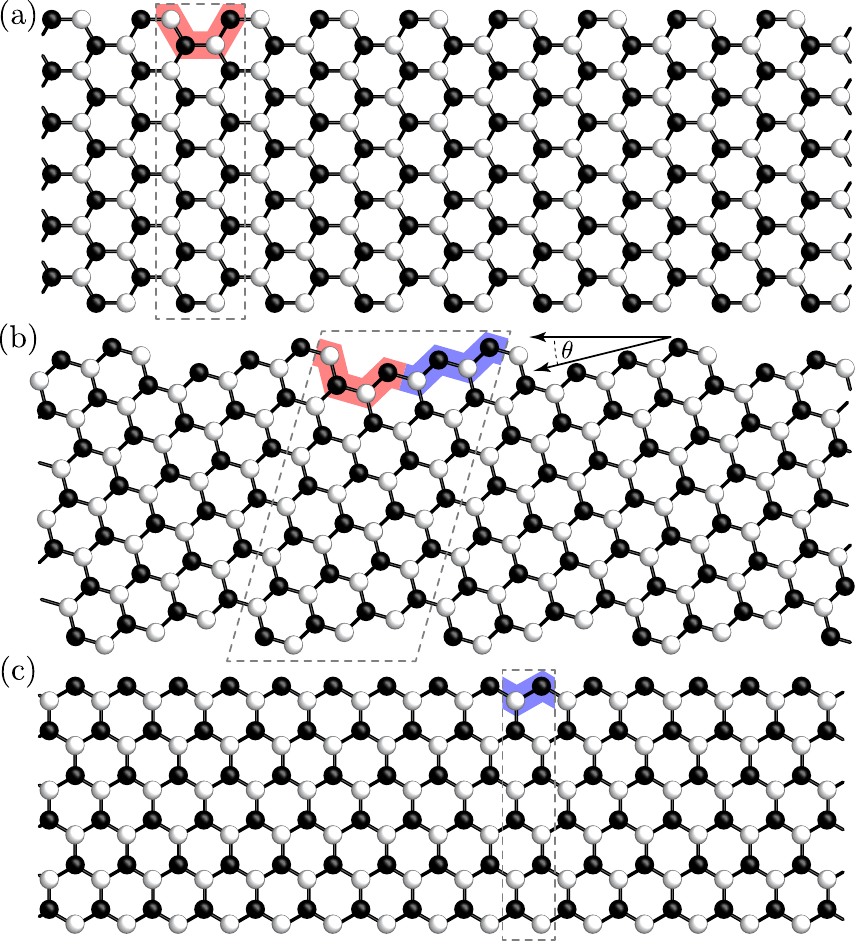}
  \caption{(Color online) (a) Armchair graphene nanoribbon of width $w=12$, denoted \mbox{$12$-aGNR}. (b) Chiral ribbon of chirality $(3,1)$ ($\theta=13.9^{\circ}$), and width $w=6$, denoted as \mbox{$6(3,1)$-cGNR}. (c) Zigzag nanoribbon of width $w=6$, denoted \mbox{$6$-zGNR}. In all cases, the two sublattices of the bipartite lattice structure are shown using light and dark spheres and the ribbons' unit cells are indicated by dashed lines. Armchair (zigzag) segments along the upper edge within the unit cell in all cases are highlighted in red (blue).
  \label{fig:lattice}}
\end{figure}
Such edge magnetism stems from the electron-electron interactions in the high density of electronic states at the Fermi level which is predominantly localized near 
the zigzag edges.\cite{Fujita96} Across the ribbon, the two edges are magnetically anti-aligned to each other~\cite{Jung09}, thus respecting Lieb's theorem of zero 
net magnetization for these bipartite lattice structures.\cite{Lieb89} Ab-initio density functional theory (DFT) calculations~\cite{Son06a,Son06b,Lee05,Pisani07}, 
as well as mean-field theory (MFT) based on the Hartree-Fock decoupling within a tight-binding Hubbard model description of zigzag GNRs predict long-ranged 
ferromagnetic 
order along the edges of zigzag GNR in the ground state.\cite{Fujita96} 
Exceedingly large magnetic correlation lengths were indeed observed in unbiased numerical studies of the edge magnetism using quantum Monte Carlo (QMC) simulations~\cite{Feldner11}, the density matrix renormalization group~\cite{Hikihara03} and exact diagonalization.\cite{Feldner10,Luitz11,Schmidt10} Steps towards a rigorous proof of the emergence of magnetic moments on zigzag edges have been put forward in Ref.~\onlinecite{Karimi12}. In the more generic case of chiral GNRs (cGNRs), characterized by the presence of both zigzag and armchair segments, the presence of low-energy edge states for edges with sufficiently long zigzag contributions has been investigated previously.\cite{Ezawa06,Akhmerov08,Wimmer10,Jaskolski11} Such ribbon geometries are in particular relevant for recent experiments that provided direct evidence for the presence of GNR edge states.\cite{Tao11,Zhang13,Li12}

Recent theoretical investigations of  chiral GNRs based on Hubbard-model MFT and ab-initio DFT studies promote the existence of spin polarized edges, similar to the 
scenario in zigzag GNRs~\cite{Yazyev11,Sun11,Barone06}. Here, we apply unbiased large-scale QMC simulations to study both the static and dynamical properties of 
chiral GNRs 
to assess the robustness of the edge magnetism with respect to the armchair segments, and compare our QMC results to self-consistent MFT calculations. 

The remainder of this paper is organized as follows: The next section introduces the nomenclature of the different GNR geometries that we analyze. After a short 
overview of the numerical method that we employ, in Sec.~III we first focus on the static magnetic properties and discuss the spin correlations for GNRs of 
different chiralities. In Sec. IV we study the local density of states (LDOS), which we calculate from the single-particle Green's function. 
We identify low-energy features in the spectral functions and local variations in the LDOS characteristic for correlated edges, which may be probed via scanning 
tunneling spectroscopy.

\section{Ribbon Geometries and Methods}

In the following, we consider several different GNR geometries, and use the standard notations to specify their edge structure: For armchair GNRs, we denote a 
ribbon with $w$ dimer lines parallel to the ribbon direction as $w$-aGNR, e.g., Fig.~\ref{fig:lattice}(a) shows a \mbox{$12$-aGNR}. A zigzag GNR ribbon with $w$ 
zigzag lines is denoted as $w$-zGNR. The \mbox{$6$-zGNR} structure is illustrated in Fig.~\ref{fig:lattice}(c). For chiral GNRs, we follow the notation in 
Ref.~\onlinecite{Yazyev11}, wherein the direction of the edge is specified by a translation vector $(n,m)$. In particular, the edge of a $(n,m)$ ribbon consists of 
a repeated unit of $m$ armchair units followed by ${n-m}$ zigzag units. The width of the chiral GNRs in terms of the number $w$ of parallel zigzag units between the 
edges is indicated by the notation \mbox{$w(n,m)$-cGNR}. The example in Fig.~\ref{fig:lattice}(b) shows a \mbox{$6(3,1)$-cGNR}. While the armchair case corresponds to $(n,m)=(1,1)$ (more specifically, a \mbox{$w(1,1)$-cGNR} is identical to a \mbox{$2w$-aGNR}) and the zigzag case to ${(n,m)=(1,0)}$, we employ the more convenient notation introduced above to highlight these symmetric edges. The chirality of a GNR edge can also be characterized by the chirality angle $\theta$, i.e., the angle between the translation vector and the closest zigzag line, which is obtained from the identity
\begin{equation}
   \sin\theta=\sqrt{\frac{3}{4}\left(\frac{m^2}{n^2+nm+m^2}\right)}\;.
\end{equation}
In chiral ribbons it spans the range $0^{\circ}<\theta<30^{\circ}$, where for zigzag edges $\theta=0^{\circ}$, and for armchair edges $\theta=30^{\circ}$. The distance between nearest neighbor sites is denoted here by $a_0$ and in the following serves as the unit for all spatial distances.
In particular, the length of the translation vector $(n,m)$, i.e. the extent of the ribbon's unit cell, is given by $a=a_0 \sqrt{n^2+nm+m^2}$.
In all cases, we employed periodic boundary conditions (PBC) along the extension of the ribbons. 

The influence of electron-electron interactions and the resulting magnetic properties of chiral GNRs in its most basic form is captured by the Hubbard model 
\begin{equation}
  H=-t \sum_{\langle i,j \rangle,\sigma} \left( c^\dagger_{i\sigma}c^{}_{j\sigma} + \mathrm{h.c.}\right) + U \sum_i n_{i\uparrow}\: n_{i\downarrow} \;.
\end{equation}
Here, $t$ denotes the hopping between nearest-neighbors $\langle i,j \rangle$ and $U$ the onsite repulsion. The operator $c^\dagger_{i\sigma}$ creates a spin-$\sigma$ fermion on site $i$ and the local density is defined by $n_{i\sigma}=c^\dagger_{i\sigma}c^{}_{i\sigma}$. Estimates of the local Coulomb repulsion in graphene-based materials put the ratio $U/t$ in the region near unity~\cite{Yazyev11,Wehling11,Schueler13}. In the following, we thus consider the weak coupling regime ${U/t\leq 2}$, where bulk magnetic order is absent~\cite{Sorella92,Meng10,Sorella12}.
Most of the simulation results are shown for ${U/t=2}$, where the magnetic properties are more pronounced and thus may be robustly detected within finite size QMC simulations.
Furthermore, we consider here the case of half-filling, where the total number of electrons equals the number of lattice sites. In this case, sign-problem free QMC simulations of the Hamiltonian $H$ can be performed on all the above introduced topologies.

We use a projective determinantal QMC method, which allows to extract ground state expectation values of an arbitrary observable $O$  (we set $\hbar=1$)
\begin{equation}
 \frac{\langle \Psi_0 | O | \Psi_0\rangle}{\langle \Psi_0 | \Psi_0\rangle}=\lim_{\Theta\to\infty} \frac{\langle \Psi_\text{T} | \E^{-\Theta H} \,O\, \E^{-\Theta H}| \Psi_\text{T}\rangle}{\langle \Psi_\text{T} |\E^{-2\Theta H}| \Psi_\text{T}\rangle} \;,
\end{equation}
by projection from a trial wave function $|\Psi_\text{T}\rangle$, which is taken here to be the eigenstates of the free system ($U=0$). The projection parameter $\Theta$ is chosen sufficiently large, such that convergence to the ground state $|\Psi_0\rangle$ is guaranteed. Depending on the detailed GNR structure, values of $\Theta$ between $\Theta=60/t$ and $100/t$ are required to ensure convergence. We use a symmetric Suzuki-Trotter decomposition  with an imaginary time discretization of $\Delta\tau=0.05/t$, such that discretization errors are well below the size of the statistical errors. Furthermore, a SU(2) symmetric Hubbard-Stratonovich decoupling of the Hubbard interaction was employed in order to ensure the explicit conservation of spin rotational symmetry. Details of the projective QMC method can be found in Ref.~\onlinecite{Assaad08}. We simulate lattices of $N_\text{C}$ unit cells along the ribbon and impose PBC. For all the considered ribbons, we specify the value of $N_\text{C}$ available for the largest simulated structure in detail below. As an example, for the \mbox{$6(3,1)$-cGNR} structure shown in Fig.~\ref{fig:lattice}(b), which contains $48$ sites per unit cell, we simulated periodic ribbon rings with up to $N_\text{C}=12$ unit cells, i.e. a total of 576 sites.

In addition to the QMC simulations we solve the self-consistent mean-field equations for the magnetization, which we obtain from the Hartree-Fock decoupling of the Hubbard interaction ${n_{i\uparrow}\: n_{i\downarrow} \rightarrow n_{i\uparrow}\langle n_{i\downarrow}\rangle + \langle n_{i\uparrow}\rangle n_{i\downarrow} - \langle n_{i\uparrow}\rangle \langle n_{i\downarrow} \rangle}$,  ignoring the fluctuation term. For regular ribbon geometries, a Fourier transformation along the ribbon direction is performed.


\begin{figure}[t!]
\centering
  \includegraphics[width=\columnwidth]{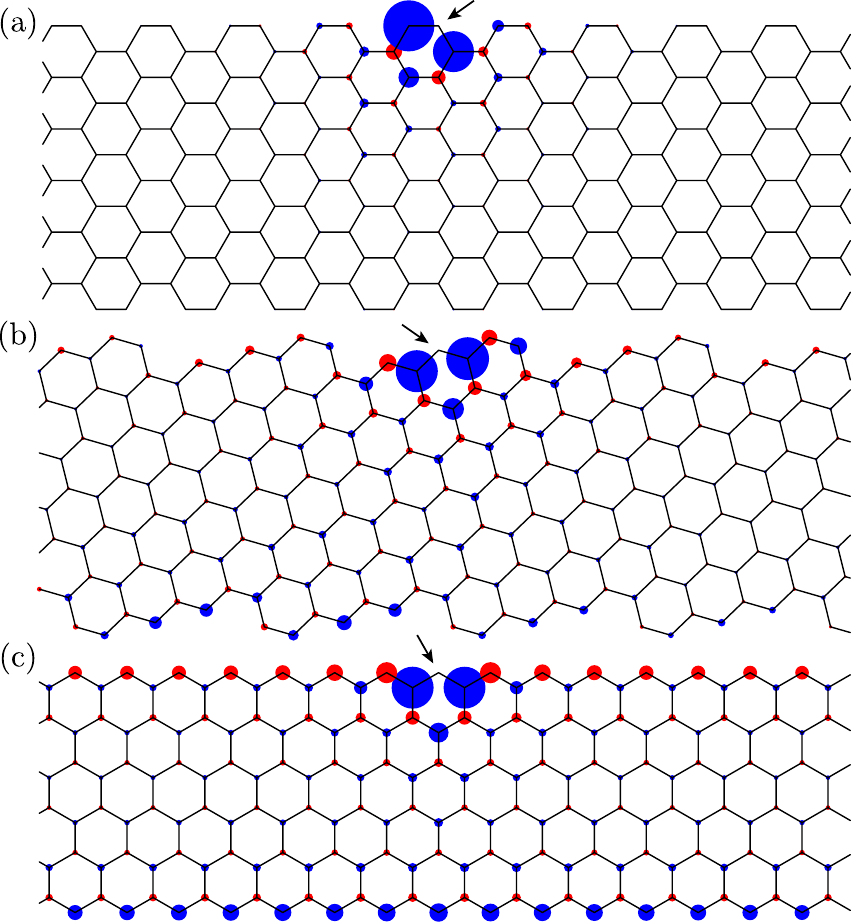}
  \caption{(Color online) 
Real-space spin correlations illustrated by disks proportional to $\langle \mathbf{S}_i \cdot\mathbf{S}_j\rangle$ across GNRs of various chiralities of width $w=6$ (corresponding to a \mbox{$12$-aGNR} in the armchair case) at $U/t=2$. In each case, the reference site is indicated by an arrow. Red (blue) circles correspond to positive (negative) values.
\label{fig:realspacecorr}}
\end{figure}

\section{Magnetic Correlations}

As a consequence of the finite density of states at the Fermi level, cGNRs have been predicted to exhibit an instability toward a spin-polarized edge  upon the introductions of electron-electron interactions. Based on MFT this edge magnetism is expected to persist essentially over the full range of chiralities, but for the near armchair limit of $\theta=30^{\circ}$.\cite{Kumazaki08,Yazyev11} In the zigzag limit, for which the edge polarization shows a maximum, exceedingly large spin correlation lengths have been confirmed by means of unbiased QMC simulations.\cite{Feldner11} In contrast, GNRs with perfect armchair edges of widths $w\neq 3k+2$ (with integer $k=0,1,2,..$) yield exponentially decaying (short range) spin correlations. This can be directly related to the semiconducting nature of such GNRs. Armchair ribbons with $w=3k+2$ in the non interacting limit are metallic and exhibit unique behavior, discussed in the following.  

Due to the  presence of repeated armchair segments in chiral GNR, the issue arises, whether the long-range spin order that results within the MFT approximation is actually robust to quantum fluctuations. To address this issue, we performed QMC simulations for GNRs of different chiralities and widths. To probe for tendencies towards edge magnetism within our QMC simulations, we measure the equal-time spin-spin correlations between sites
and in particular monitor the spin correlation function $C(r)=\langle \mathbf{S}_0 \cdot\mathbf{S}_r\rangle$ as a function of the distance $r$ along the edges of the GNRs. 

\begin{figure}[t!]
\centering
  \includegraphics[width=\columnwidth]{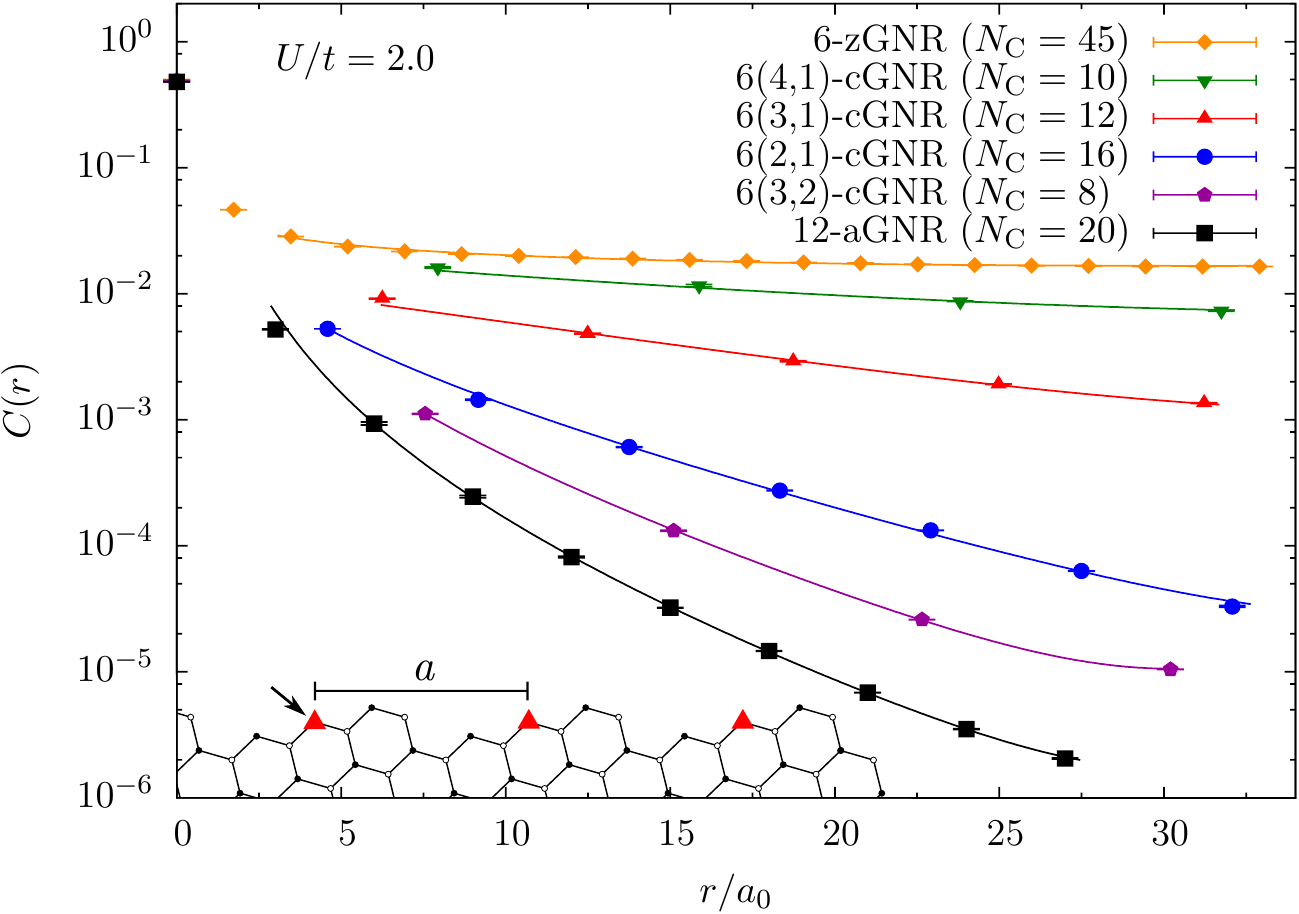}
  \caption{(Color online) 
Spin correlation function $C(r)$ along the edge of GNR of various chiralities of width $w=6$ (corresponding to a \mbox{$12$-aGNR} in the armchair case). For each case, the employed number of unit cells $N_\text{C}$ is indicated.
Lines show the fitting function $C_{\text{fit}}(r)$ for each GNR case.
The inset in the lower left part of the figure indicates the position of the reference site for the case of a (3,1)-cGNR edge (arrow), as well as the distance to the corresponding sites (triangles) within further unit cells, each  separated by a distance $a$.    
  \label{fig:spinspinw6}}
\end{figure}

The general correlation pattern along the ribbon edge, as well as between the two edges and into the bulk, is illustrated in the real-space representations of the correlations in Fig.~\ref{fig:realspacecorr} for three characteristic cases, also considered in Fig.~\ref{fig:spinspinw6}. The reference sites are indicated by arrows. The extremely short range correlations in the armchair GNR in Fig.~\ref{fig:realspacecorr}(a) steadily increase from the 6(3,1)-cGNR in Fig.~\ref{fig:realspacecorr}(b) to the zigzag GNR in Fig.~\ref{fig:realspacecorr}(c). In all cases the correlations decay quickly into the bulk and are anti-aligned to each other on opposite ribbon edges, respecting the sublattice structure. Like in zigzag GNRs, the correlations in the chiral GNR are largest along the zigzag segments interrupted by armchair elements where correlations are suppressed. Compared to the zigzag case, the correlations into the bulk decay quicker and are strongest along the direction towards the closest opposite edge.

\begin{figure}[t!]
\centering
  \includegraphics[width=\columnwidth]{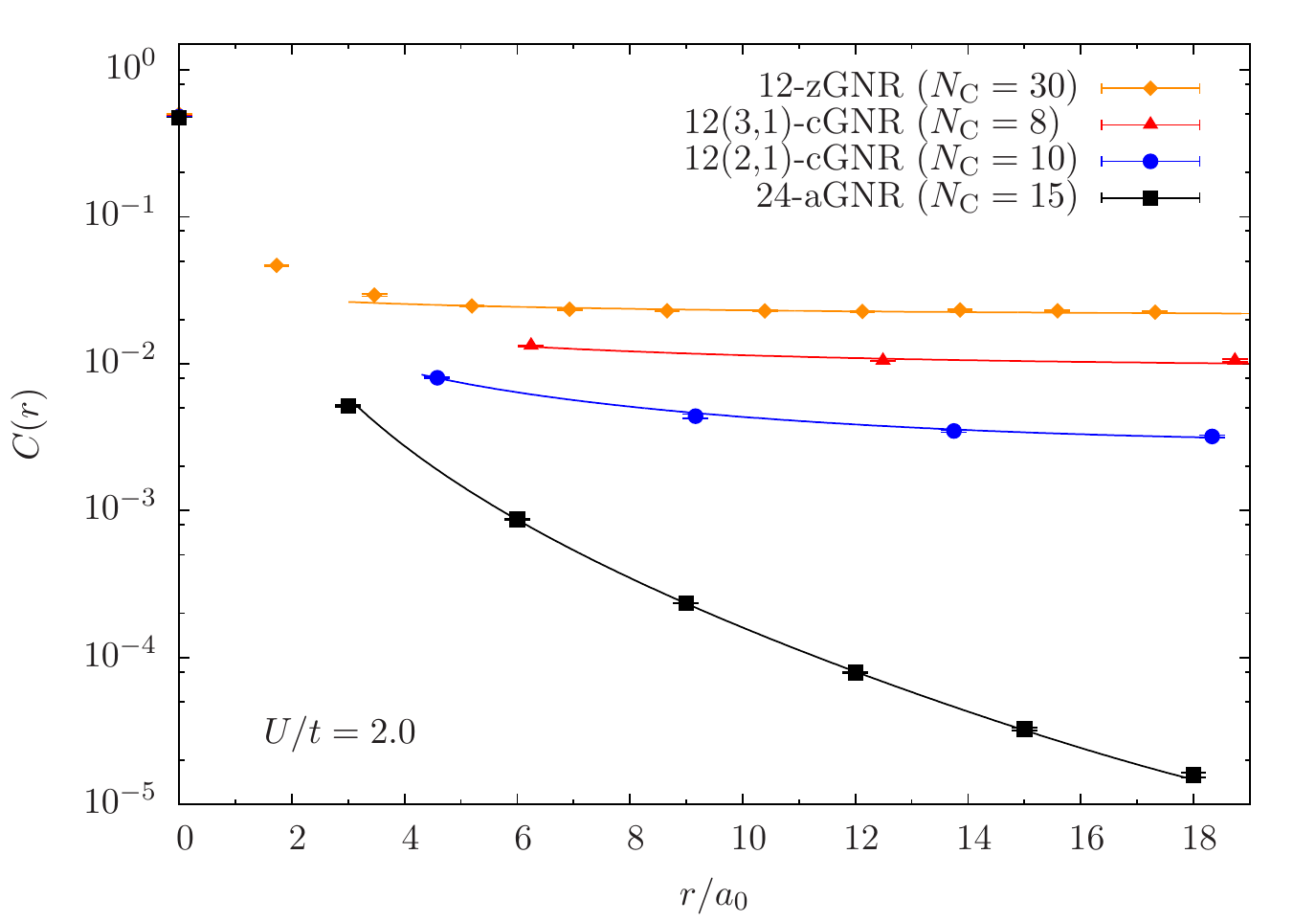}
  \caption{(Color online) 
Spin correlation function $C(r)$ along the edge of GNR of various chiralities of width $w=12$ (corresponding to a $24$-aGNR in the armchair case). For each case, the employed number of unit cells $N_\text{C}$ is indicated.
Lines show the fitting function $C_{\text{fit}}(r)$ for each GNR case. 
  \label{fig:spinspinw12}}
\end{figure}

In Fig.~\ref{fig:spinspinw6}, we show results for the correlation function $C(r)$ as a function of the distance $r$ from a given reference site for GNRs of different chiralities and width $w=6$ (in the armchair case, this corresponds to a \mbox{$12$-aGNR} in standard notation).These results were obtained on the longest ribbons of each chirality type accessible in our QMC simulations. The correlations are shown among edge sites of the same sublattice with respect to a reference site centrally located within the zigzag segment for chiral GNRs.  Consecutive data points correspond in each case to a spatial separation of $a$. For better comparison among the different cases, we  plot all distances $r$ in units of $a_0$, the nearest neighbor distance on the graphene lattice.
The inset of Fig.~\ref{fig:spinspinw6}  illustrates this procedure for the case of a 3,1)-cGNR.
Note that for armchair and zigzag GNRs, all outer edge sites provide equivalent reference points.

For this intermediate ribbon width ($w=6$), we observe the clear tendency towards quasi long-range spin correlations. Starting from the nearly exponentially decaying correlations in the aGNR, they become increasingly dominated by power law behavior with increasing length of the zigzag segments, before essentially long-range spin correlations prevail. We estimate the spin correlation length $\xi$ for the various chiralities by fitting a function that combines both, exponential and power-law behavior:
\begin{equation}
 C_{\text{fit}}(r)\propto r^{-\eta} \E^{-r/\xi} + (L-r)^{-\eta} \E^{-(L-r)/\xi}\;.
\end{equation}
Here the second term accounts for the PBC, with $L$ denoting the linear extent of the ribbon along its edge.\cite{Feldner11} For these ribbons, the fitted values of the exponent $\eta$ range between $0.2$ and $1.9$ and we concentrate on discussing the correlation length $\xi$ in the following. For the 12-aGNR armchair ribbon we extract an expectedly short correlation length of $\xi=3.2(2)a_0$. The 6(2,1)-cGNR, 6(3,1)-cGNR and 6(4,1)-cGNR show a  $\xi/a_0$ which steadily increases from $6.5(5)$ via $13.6(6)$ to $26(2)$, respectively (numbers in brackets quantify the uncertainty in the last digit). The resulting fitting functions $C_{\text{fit}}(r)$ are shown along with the QMC data in Fig.~\ref{fig:spinspinw6}.
In the 6(3,2)-cGNR, with two armchair segments separating the zigzag regions, we find a stronger exponential decay with $\xi=5.7(1)a_0$. Similar to the armchair case this ribbon lacks spin polarized edges even in MFT below $U/t=1.94$ (not shown). In the zigzag ribbon, the spin correlation length extracted from the fit exceeds the system size even on this largest available system, which is in accordance with previous QMC results.\cite{Feldner11} 
\begin{figure}[t!]
\centering
  \includegraphics[width=\columnwidth]{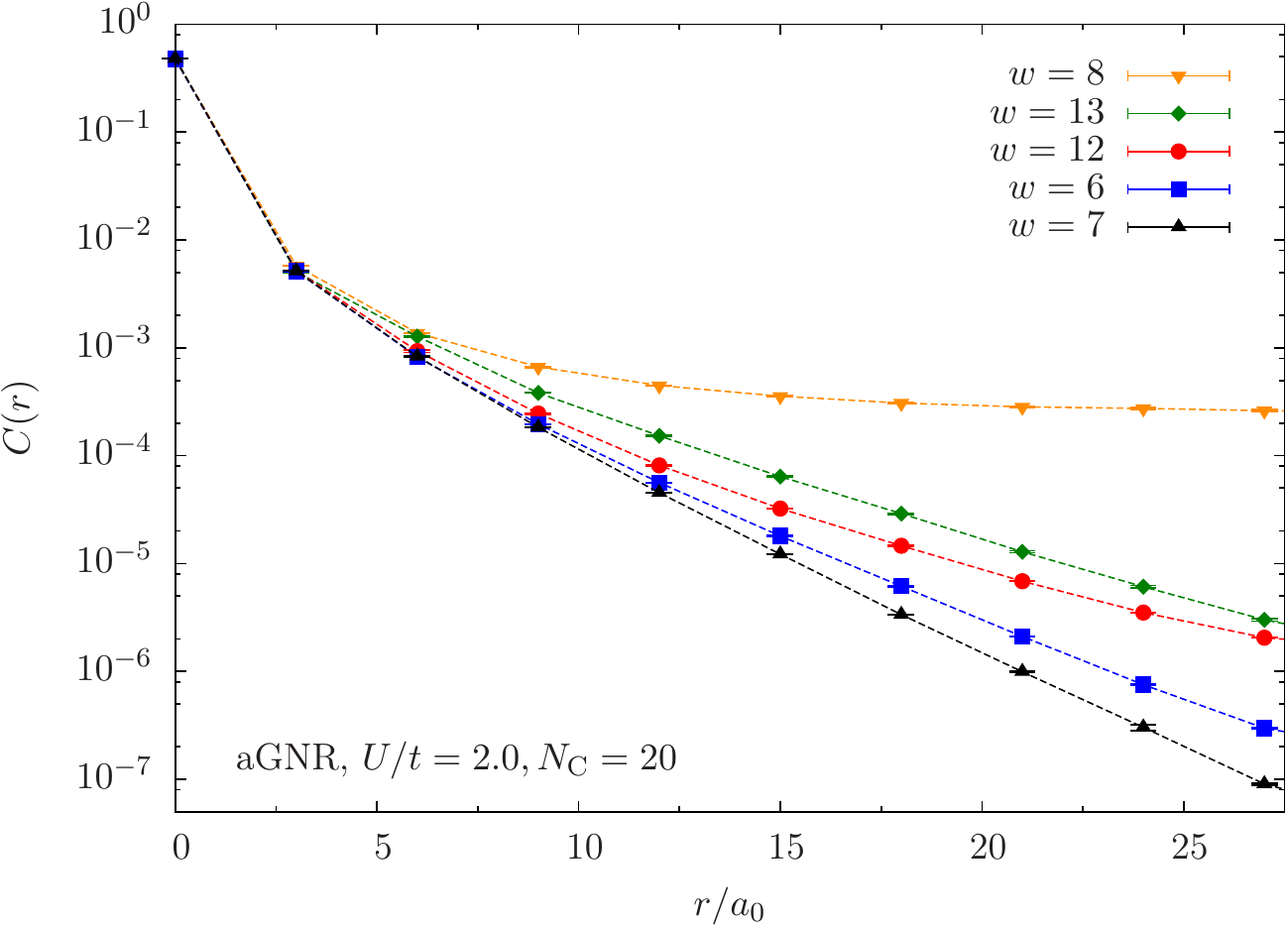}
  \caption{(Color online) 
Spin correlation function $C(r)$ along the edge of aGNRs of different widths $w$. Lines in this figure are guides to the eye. 
  \label{fig:spinspinagnr}}
\end{figure}

The generally rapid decay of the spin correlations along armchair edges is presented in Fig.~\ref{fig:spinspinagnr}, which shows results for armchair GNRs of 
different widths. A noticeable exception from this trend is provided by the \mbox{8-aGNR}: In contrast to other armchair GNRs, where $\xi\lesssim 6 a_0$, we extract 
much stronger correlations. This behavior traces back to the metallic nature of the \mbox{8-aGNR} in the $U=0$ tight binding limit, which is among the series 
$w=3k+2$ (with integer $k=0,1,2,...$).  It is not related to the emergence of edge magnetism, in particular since the low-energy states in this case are not edge-localized. 
Instead, the behavior here is similar to the Hubbard model on the one-dimensional chain (to which indeed the armchair GNR degrades for $w=2$), for which $U>0$ triggers an instability towards a quasi-long-ranged-ordered bulk antiferromagnetic spin density wave state.

\begin{figure}[t!]
\centering
  \includegraphics[width=\columnwidth]{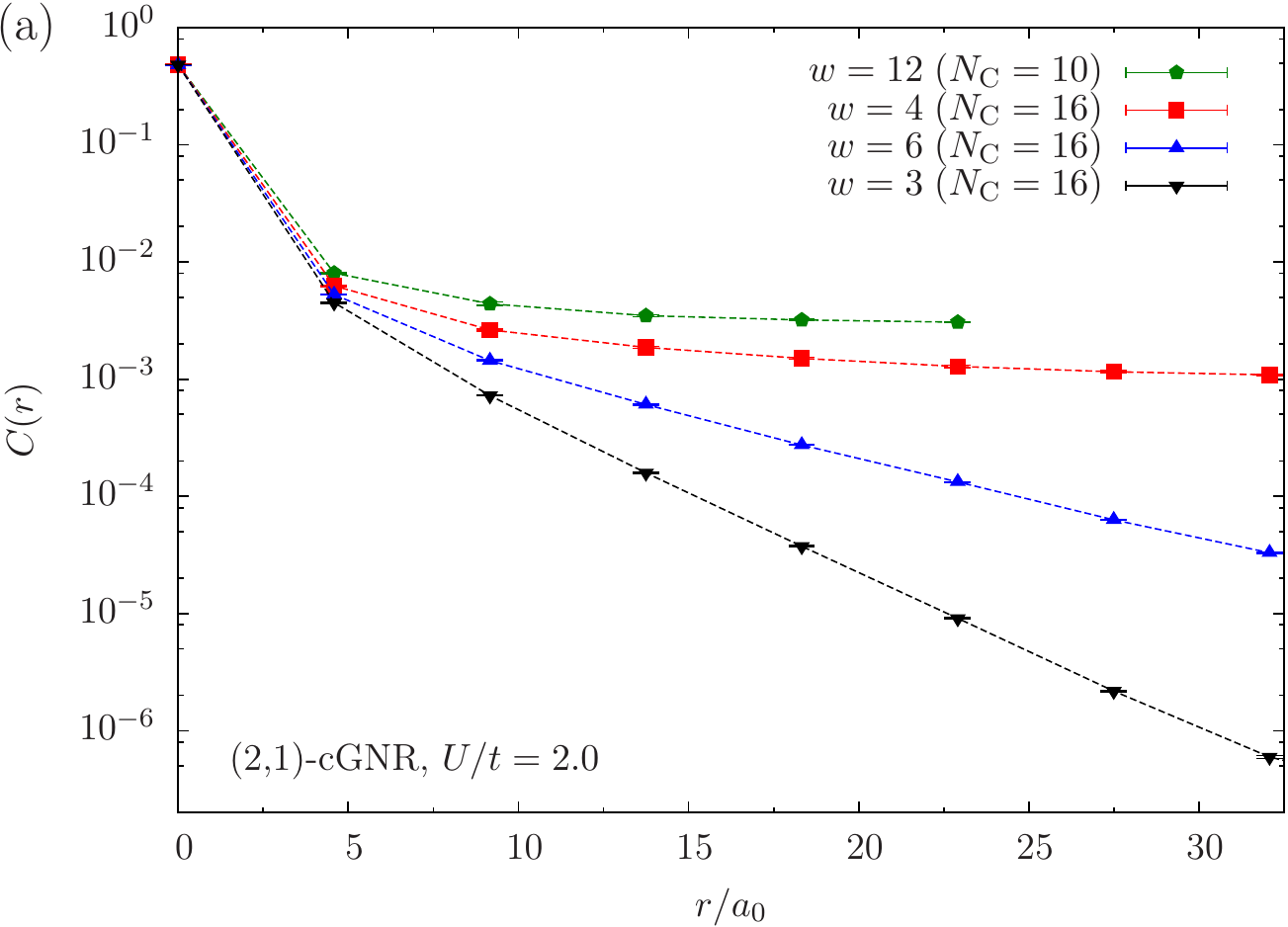}\\
  \includegraphics[width=\columnwidth]{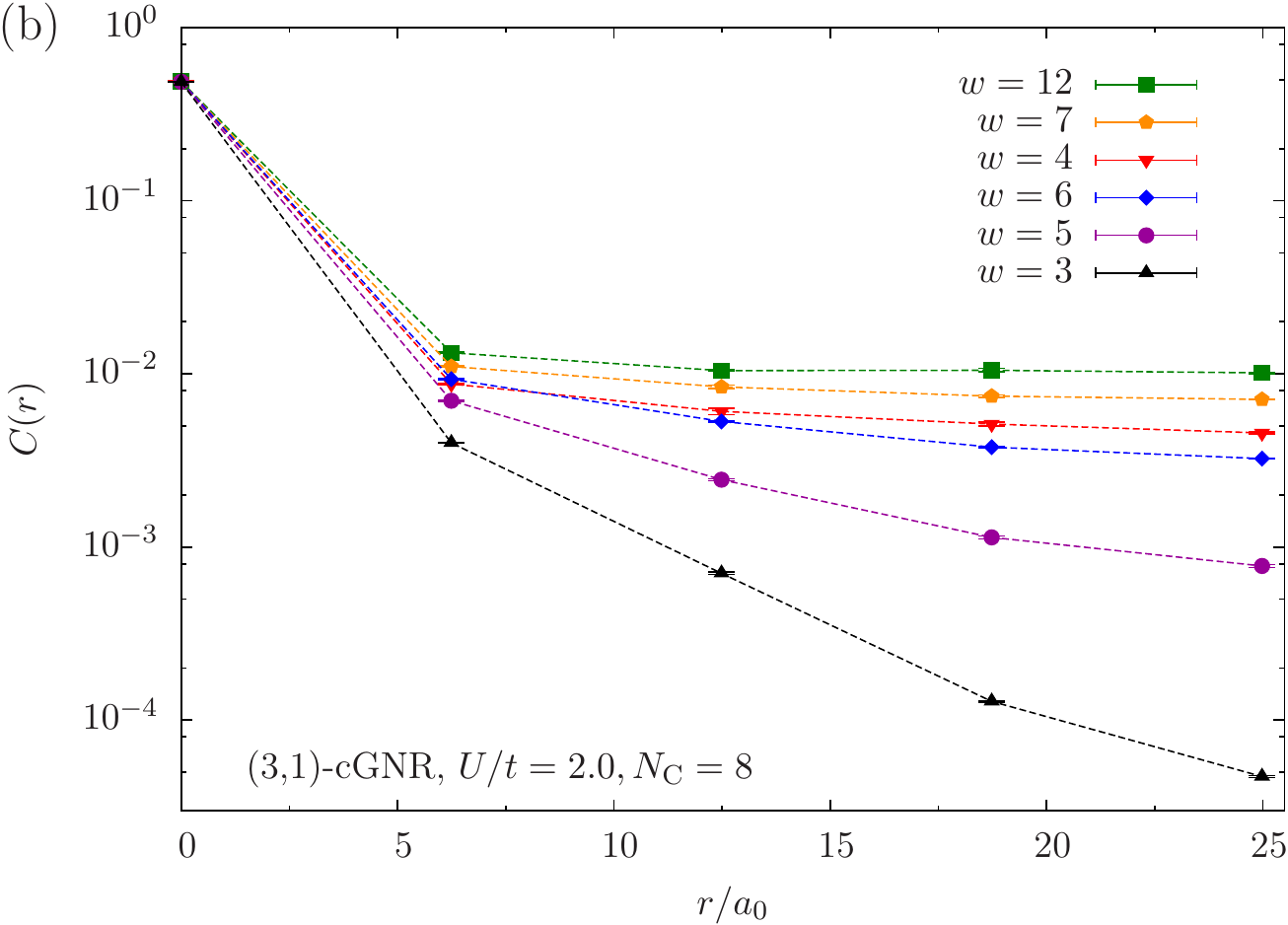}
  \caption{(Color online) 
Spin correlation function $C(r)$ along the edge of (a) (2,1)-cGNR and (b) (3,1)-cGNR of different widths. Lines in this figure are guides to the eye. 
  \label{fig:spinspin23cgnr}}
\end{figure}

\begin{figure}[t!]
\centering
  \includegraphics[width=\columnwidth]{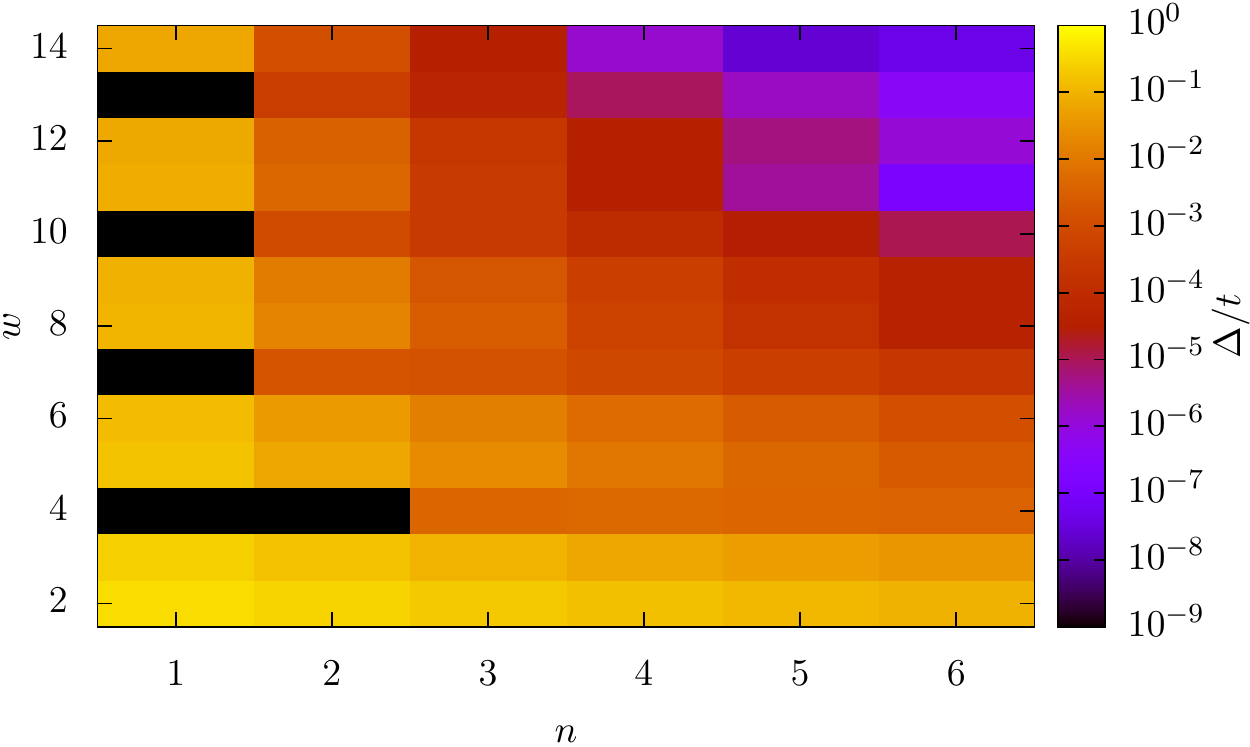}
  \caption{(Color online) 
Map of the single particle excitation gap for the $U=0$ tight-binding limit for $w(n,1)$-cGNR of varying $n$ (i.e. chirality) and width $w$. 
  \label{fig:gapmap}}
\end{figure}

In increasingly wider chiral GNRs,  the tendency towards edge magnetism becomes more established. This can be seen by comparing the results in Fig.~\ref{fig:spinspinw6} for $w=6$ to those in Fig.~\ref{fig:spinspinw12}, which provides data on different ribbons twice the width ($w=12$ for cGNR, corresponding to $w=24$ for the armchair case). While the $w=24$ armchair ribbon still shows a fast decay of the correlations, resulting in a similar correlation length as for $w=12$, we find significantly enhanced correlations for the \mbox{12(2,1)-cGNR}, and the \mbox{12(3,1)-cGNR}. Their estimates for the $\xi$ exceed  the available system sizes. Enhanced correlations for specific widths can also be observed in chiral ribbons. For example, they are present along the \mbox{4(2,1)-cGNR} and the \mbox{4(3,1)-cGNR} (cf. Fig.~\ref{fig:spinspin23cgnr}).  Compared to neighboring values of $w$, such enhanced correlations at $w=4$ may be associated with a vanishing (for the \mbox{4(2,1)-cGNR}) or very low (for the \mbox{4(3,1)-cGNR}) single particle energy gap in the tight binding limit. Similar behavior we thus expect to also be observed for other specific widths of chiral ribbons, and is reflected by the horizontal stripes for $w=4,7,10,13$ in the single particle gap map for $U=0$ shown in Fig.~\ref{fig:gapmap}. These ribbons are special examples of the metallic or almost metallic (MAM) points identified in Ref.~\onlinecite{Ezawa06}. 

\section{Dynamic spectral functions}

\begin{figure*}
\centering
  \includegraphics[width=2\columnwidth]{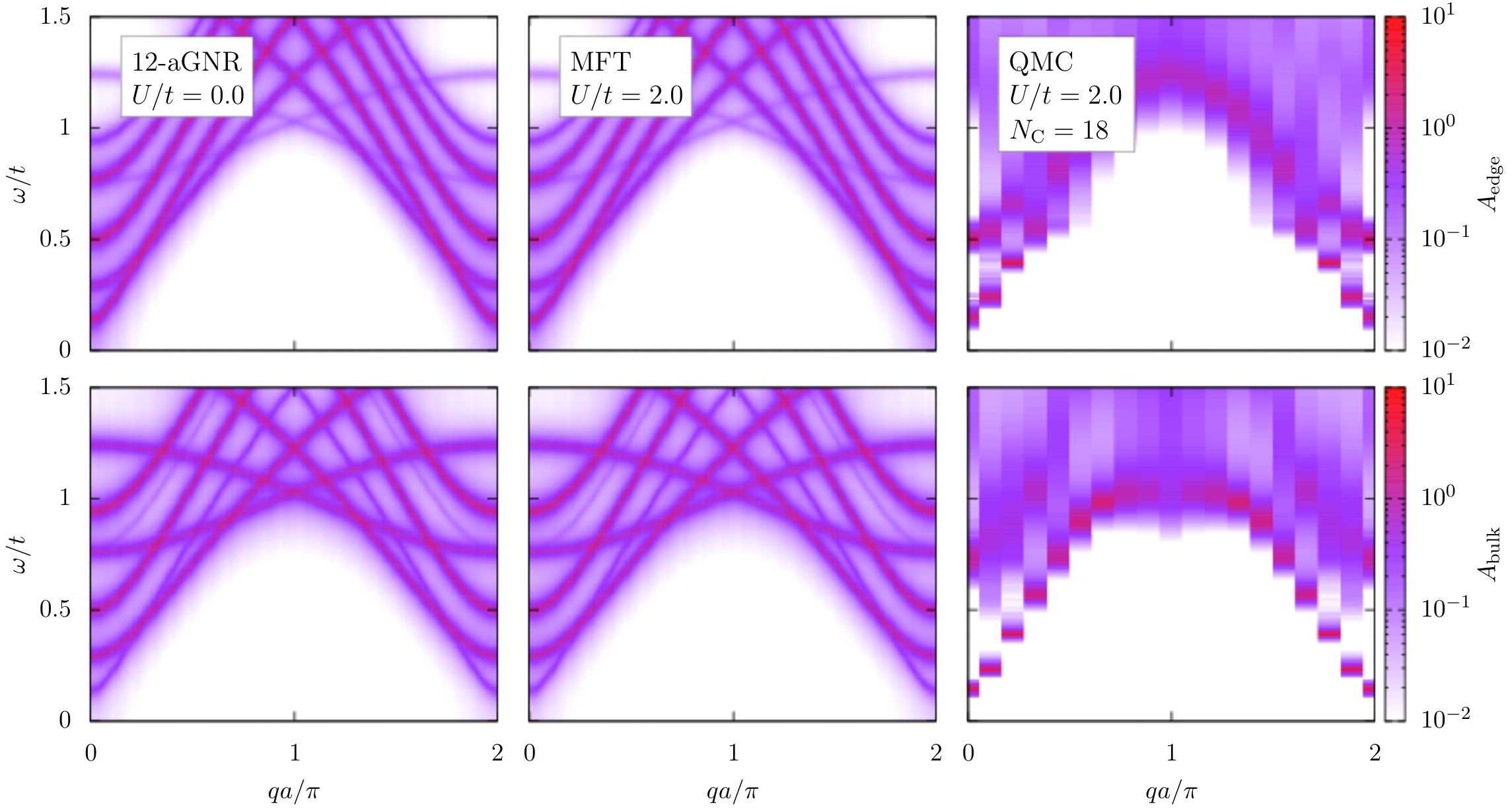}
  \caption{(Color online) Momentum-resolved single particle spectral function for a 12-aGNR for edge sites (top) and bulk sites (bottom), for $U=0$ (left), and from MFT (middle) and QMC simulations (right) for $U/t=2$.   
  \label{fig:Aqa}}
\end{figure*}

\begin{figure*}
\centering
  \includegraphics[width=2\columnwidth]{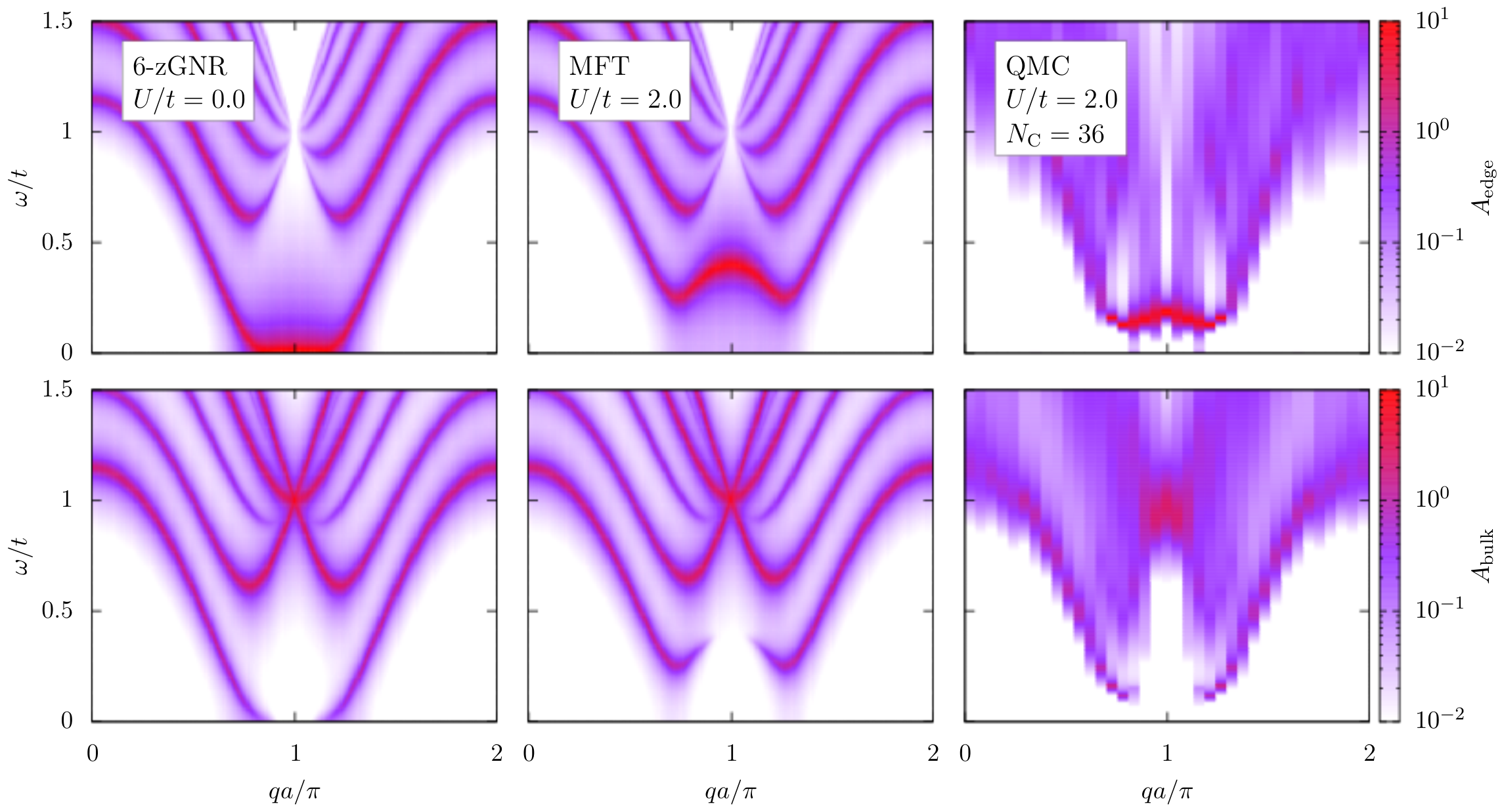}
  \caption{(Color online) Momentum-resolved single particle spectral function for a 6-zGNR for edge sites (top) and bulk sites (bottom), for $U=0$ (left), and from MFT (middle) and QMC Simulations (right) for $U/t=2$.   
  \label{fig:Aqz}}
\end{figure*}

The equal-time spin-spin correlations, examined in Sec. III, allow to quantify the strength of the magnetic correlations along the GNRs.
We next consider the dynamical single-particle spectral function, which is particularly important, as it may be directly compared with 
scanning tunneling spectroscopy experiments. The spin resolved local and momentum-resolved spectral functions are given at frequency $\omega$ by
\begin{equation}
 A_{i\sigma}(\omega)=\sum_n|\langle \Psi_0|c_{i\sigma} | \Psi_n \rangle |^2\; \delta(\omega+E_0-E_n)\;,
\end{equation}
and
\begin{equation}
 A_{\alpha\sigma}(q,\omega)=\sum_n|\langle \Psi_0|c_{\alpha q\sigma} | \Psi_n \rangle |^2\; \delta(\omega+E_0-E_n)\;,
\end{equation}
respectively, 
where $c_{\alpha q\sigma}=\frac{1}{\sqrt{N_\text{C}}} \sum_{j=1}^{N_\text{C}} \E^{\I q a j} c_{i(\alpha, j)\sigma}$, with $i(\alpha, j)$ denoting the site index of the lattice site located at position $\alpha$ within the $j$-th unit cell, and $n$ enumerates the full set of eigenstates $| \Psi_n \rangle $ with energy $E_n$ of the MFT Hamiltonian. Because of particle-hole symmetry, we need to consider only the range of positive $\omega \geq 0$. Within MFT, broken spin rotational symmetry leads to distinct spectral functions for both spin orientations, which we average to $A_i(\omega)=\frac{1}{2}[A_{i\uparrow}(\omega)+A_{i\downarrow}(\omega)]$. In the QMC simulations, SU(2) symmetry is preserved and therefore $A_i(\omega)=A_{i\uparrow}(\omega)=A_{i\downarrow}(\omega)$. The Dirac $\delta$-functions in the above formula were subjected to a Lorentzian broadening of $\Delta\omega=0.02t$ within the MFT calculations.  Within the QMC method we cannot directly access real time correlation functions, and thus measure the momentum-resolved Green's function $G_\alpha(q,\tau)=\frac{1}{2}\sum_\sigma\langle \Psi_0 | c^{}_{\alpha q\sigma}(\tau) c^{\dagger}_{\alpha q\sigma}(0)| \Psi_0 \rangle$ in imaginary time $\tau$, instead.\cite{Feldbacher01} The spectral function on the real frequency axis $\omega$ may be obtained by the inversion of
\begin{equation}
   G_\alpha(q,\tau)=\int \: d\omega \: \E^{-\tau\omega} \:A_\alpha(q,\omega)\;,
\end{equation}
by means of the stochastic analytic continuation method.\cite{Beach04} The local spectral function $A_i(\omega)$ is then calculated from the corresponding momentum-resolved spectral function $A_\alpha(q,\omega)$ by integrating over the momentum $q$ along the ribbon direction. This procedure ensures that features in the local spectral function related to different momenta $q$ are preserved by the analytic continuation. In the following, we consider the spectral functions for sites along the ribbon edge, denoted by $A_\text{edge}(q,\omega)$ and $A_\text{edge}(\omega)$, and within the center (bulk) of the ribbon denoted by $A_\text{bulk}(q,\omega)$, and $A_\text{bulk}(\omega)$, respectively. For chiral ribbons the edge site refers to a site in the center of a zigzag segment, unless noted otherwise. 

\begin{figure*}
\centering
  \includegraphics[width=2\columnwidth]{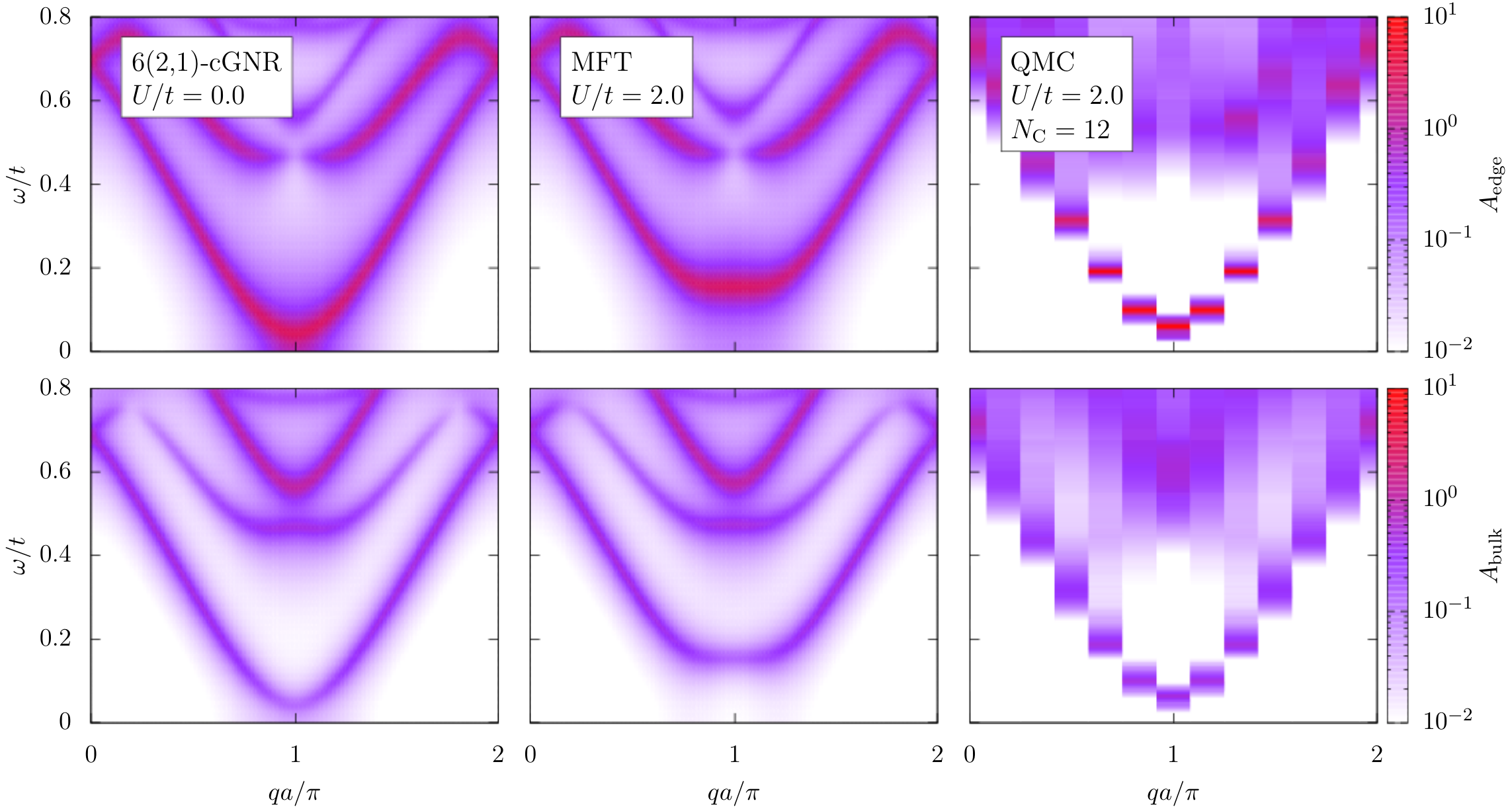}
  \caption{(Color online) Momentum-resolved single particle spectral function for a \mbox{6(2,1)-cGNR} for edge sites (top) and bulk sites (bottom), for $U=0$ (left), and from MFT (middle) and QMC simulations (right) for $U/t=2$.   
  \label{fig:Aq21}}
\end{figure*}

\begin{figure*}
\centering
  \includegraphics[width=2\columnwidth]{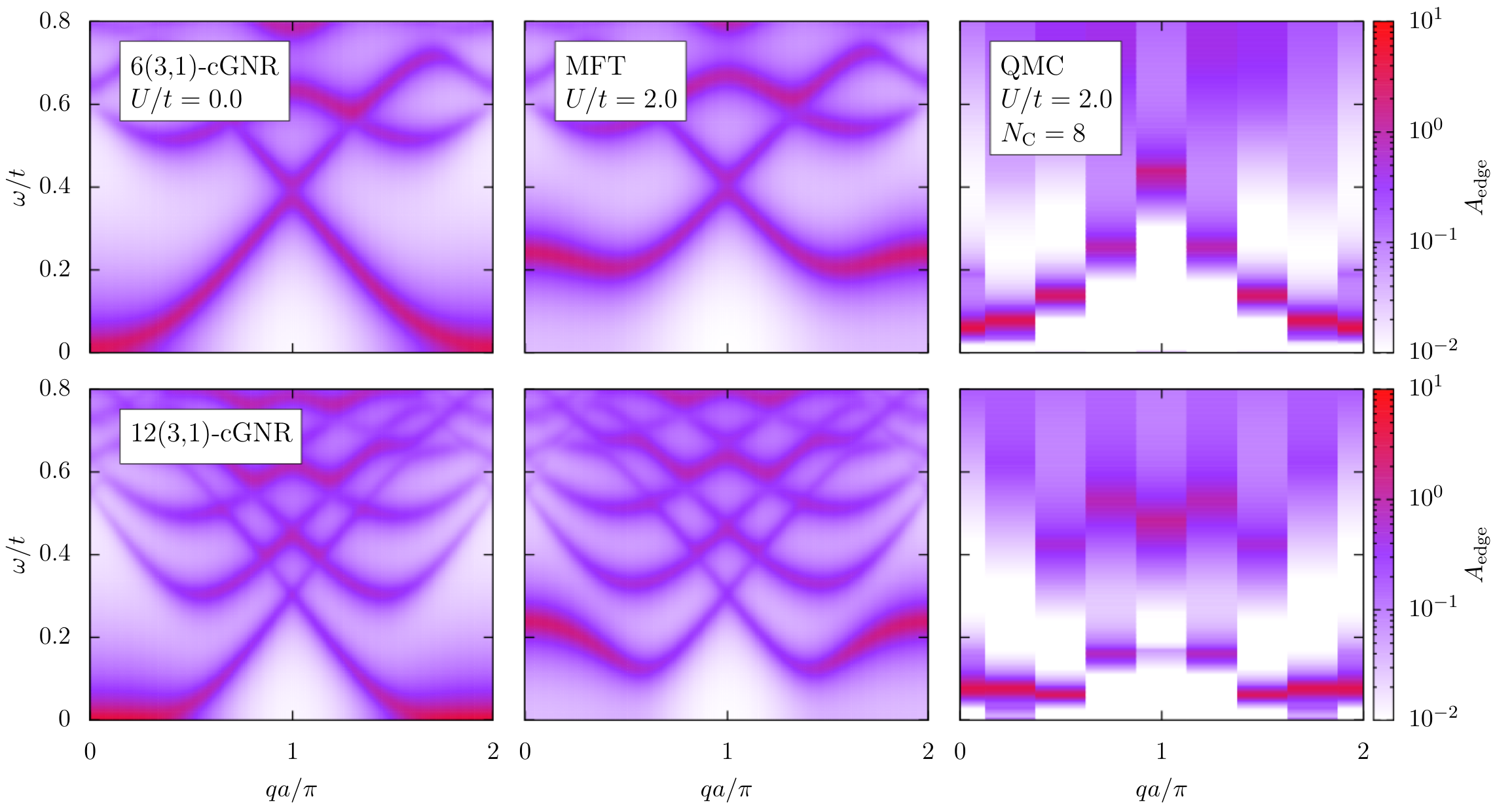}
  \caption{(Color online) Momentum-resolved single particle spectral function along the edge sites for a \mbox{6(3,1)-cGNR} (top) and a \mbox{12(3,1)-cGNR} (bottom), for $U=0$ (left), and from MFT (middle) and QMC simulations (right) for $U/t=2$.   
  \label{fig:Aq31}}
\end{figure*}

In order to discuss the effects of electron interactions on the LDOS, it is instructive to first examine the momentum-resolved local spectral functions for GNRs of different chiralities. For this purpose, we show in \mbox{Figs.~\ref{fig:Aqa}--\ref{fig:Aq31}} QMC results for $A_\text{edge}(q,\omega)$ and $A_\text{bulk}(q,\omega)$, along with tight-binding ($U=0$) as well as with MFT results.

Starting from the spectrum of the \mbox{12-aGNR} (cf. Fig.~\ref{fig:Aqa}), the $U=0$ tight-binding and the $U/t=2$ MFT results are in fact identical. This is due of the absence of a finite spin polarization in the mean-field solution at this coupling strength.  Upon comparing to the QMC spectral functions, one has to account for  inherent resolution limitations imposed by the analytic continuation of the imaginary-time QMC data. While low energy features are usually well defined, the analytic continuation fails to resolve or broadens individual features at high energies. Here, we indeed find that the low energy region of the QMC spectral function compares well to the MFT results, revealing the same low energy characteristics. Both exhibit a similar excitation gap of $\Delta/t\approx 0.14$. We hence find no significant difference to the non-interacting case.

\begin{figure}[t!]
\centering
  \includegraphics[width=\columnwidth]{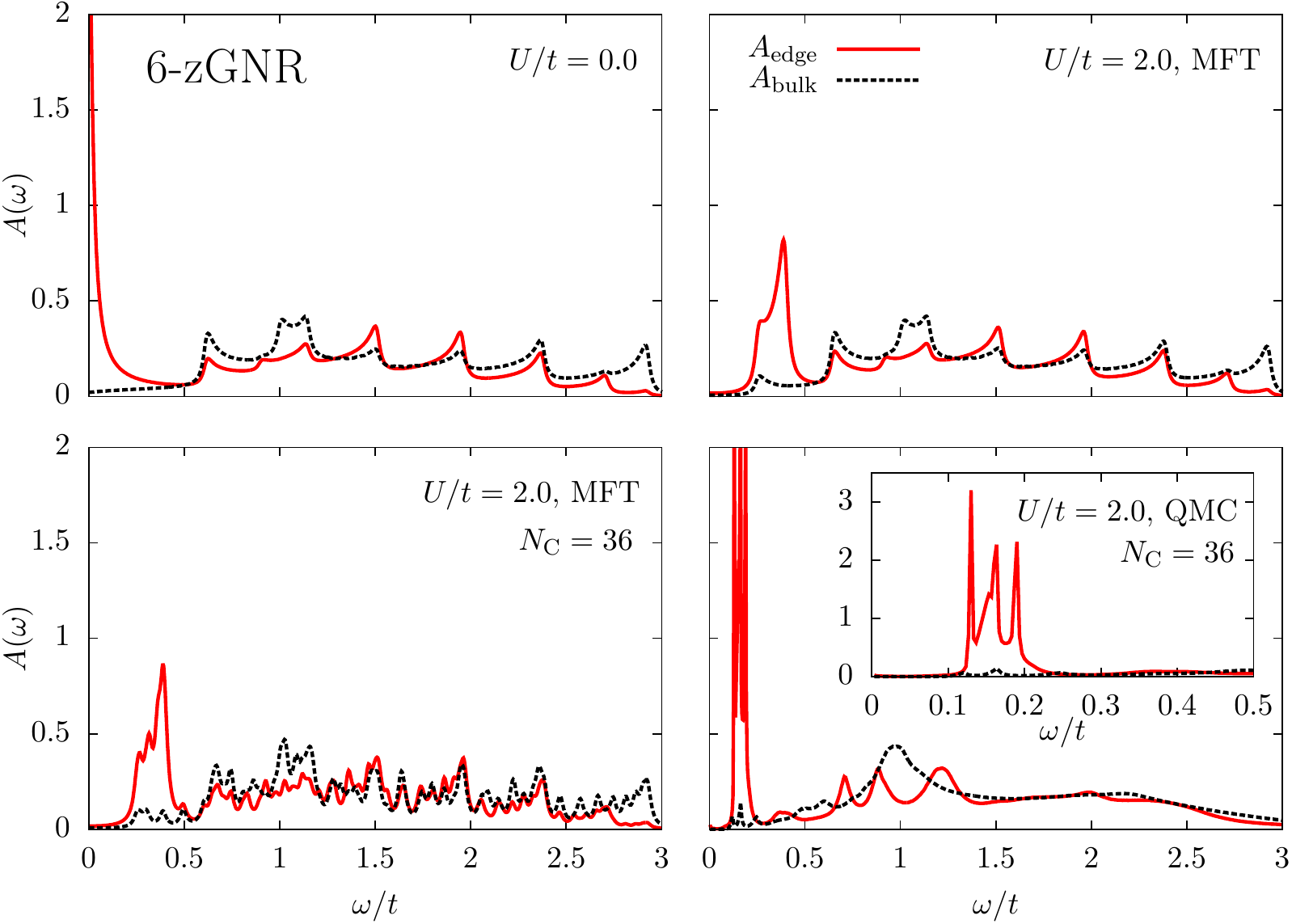}
  \caption{(Color online) Local single particle spectral function for a 6-zGNR for edge sites and bulk sites.  The inset focuses in on the QMC data at low energies.  
  \label{fig:zgnr_local}}
\end{figure}

In contrast, in the zigzag GNR results shown in Fig.~\ref{fig:Aqz}, we observe distinct spectral features related to edge magnetism at finite $U$, as already reported previously~\cite{Feldner11,MengThesis,FeldnerThesis}. At $U=0$, the spectral function traces the low energy edge states with a flat dispersion within the momentum range ${2/3\leq  q a/\pi \leq 4/3}$. Both MFT and QMC results show the increase of the gap,  related to the onset of the edge magnetism and the characteristic bending of the low-energy edge band with a maximum intensity at ${q=\pi/a}$. This pronounced low-energy band, located along the ribbon edge, results in a distinctive low energy peak at $\omega_\text{max}\approx 0.2t$ in the local spectral function $A_\text{edge}(\omega)$, i.e. the LDOS,  shown for the \mbox{6-zGNR} in Fig.~\ref{fig:zgnr_local}, while being absent in $A_\text{bulk}(\omega)$. As noted previously,  MFT overestimates the gap energy scales at $U/t=2$ by about a factor of two~\cite{Feldner11,MengThesis,FeldnerThesis}, in accordance with the results in Fig.~\ref{fig:zgnr_local}. The integrated weight of the QMC low energy peak in $A_\text{edge}(\omega)$ is robustly reproduced by  MFT, with only about ten percent deviations among the two approaches.  The lower left panel of Fig.~\ref{fig:zgnr_local} shows MFT results for a finite zigzag GNR with $N_\text{C}=36$ unit cells, i.e. for the same finite system as employed in the QMC simulations. This calculation reproduces the splitting of the low energy band into three separate peaks seen in the QMC data due to finite size effects. 

\begin{figure}[t!]
\centering
  \includegraphics[width=\columnwidth]{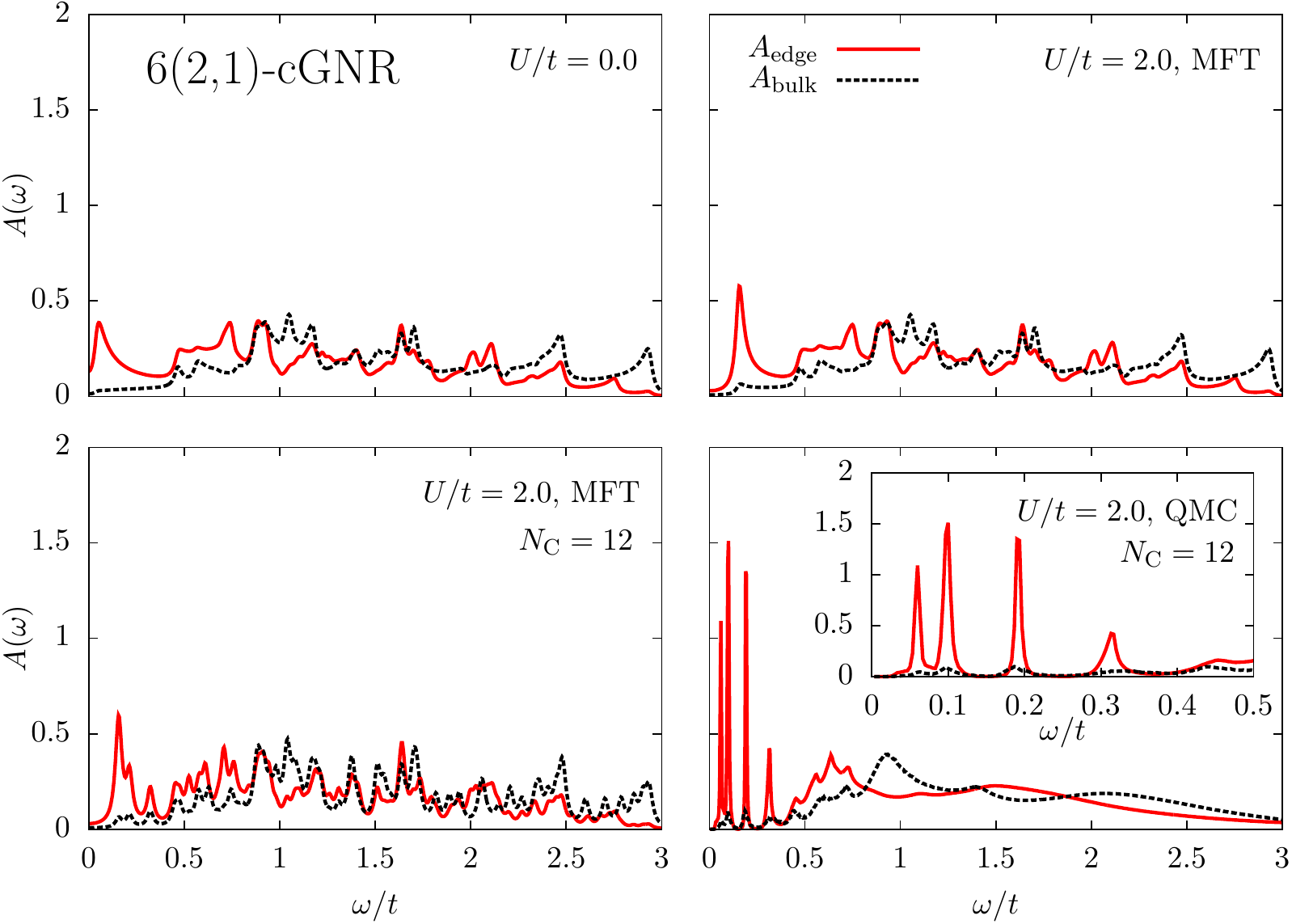}
  \caption{(Color online) Local single particle spectral function for a 6(2,1)-cGNR for edge sites and bulk sites.   The inset focuses in on the QMC data at low energies.
  \label{fig:2-1cgnr_local}}
\end{figure}

After having discussed the cases of symmetric GNR edges, we next consider the spectral functions for chiral GNRs. Figure~\ref{fig:Aq21} shows the results for the \mbox{6(2,1)-cGNR}. In the tight-binding limit, one identifies in $A_\text{edge}(q,\omega)$ a dispersing low-energy band of edge states, which for wider ribbons eventually transforms into a flat band near $q=\pi/a$, thus providing a finite DOS of edge states at the Fermi energy. These low energy states of the \mbox{6(2,1)-cGNR} extend further into the bulk compared to the pure zigzag case, which relates to the incomplete suppression of the spectral weight near $q=\pi/a$ in $A_\text{bulk}(q,\omega)$ (compare to the zigzag case). Finite interactions lead to a sizable single particle gap within MFT, with a maximum of the spectral weight centered around $q=\pi/a$ in $A_\text{edge}(q,\omega)$. Like for the zigzag case, this results in a prominent low-energy peak in the LDOS at the ribbon edge, as seen for $A_\text{edge}(\omega)$ in Fig.~\ref{fig:2-1cgnr_local}. The spectral function obtained from QMC simulations, shown in Fig.~\ref{fig:Aq21}, similarly exhibits a low energy mode predominately confined to the GNR edge, however with a significantly smaller gap. In the LDOS, this leads to the pronounced low-energy spectral weight in $A_\text{edge}(\omega)$ shown in Fig.~\ref{fig:2-1cgnr_local}. As for the zigzag case, the QMC result for the spectral function exhibits a discrete set of isolated peaks at low energies; these finite size effects are again reproduced within MFT by considering the same finite size system as in the QMC simulations (cf. the lower left panel in Fig.~\ref{fig:2-1cgnr_local}). While MFT overestimates the peak position $\omega_\text{max}$, like in the zigzag case, the integrated spectral weight within the low-energy regime in $A_\text{edge}(\omega)$
nevertheless agrees up to about a ten percent deviation among the two methods. 

\begin{figure}[t!]
\centering
  \includegraphics[width=\columnwidth]{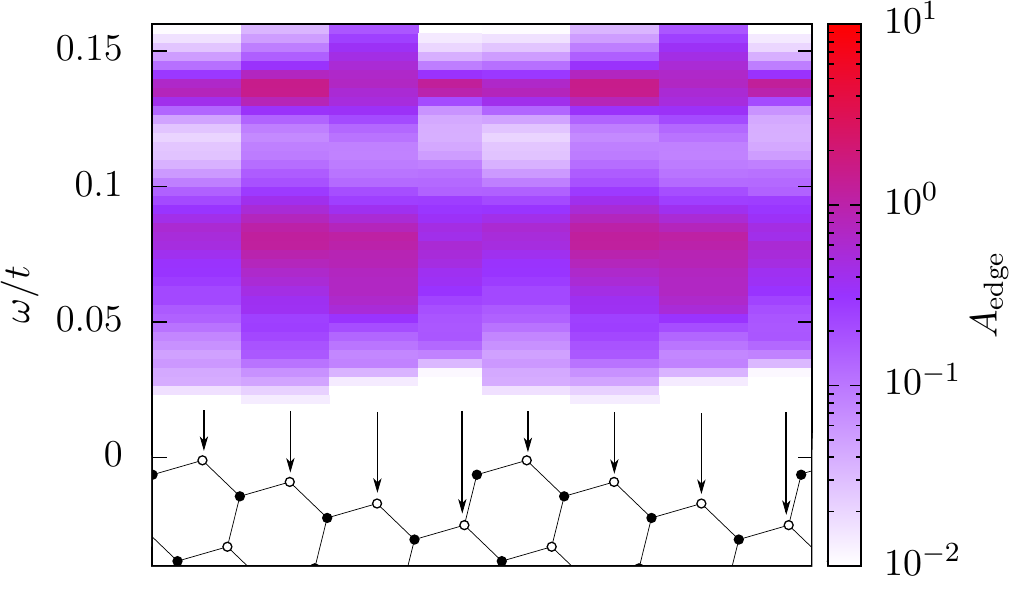}
  \caption{(Color online) Local single particle spectral function at low energies for a \mbox{6(3,1)-cGNR} from QMC simulations along the ribbon edge (shown on the bottom), within the majority sublattice, as indicated by the arrows. 
  \label{fig:edge3-1local}}
\end{figure}

For the \mbox{6(3,1)-cGNR} and the \mbox{12(3,1)-cGNR}, results for the momentum-resolved spectral functions on the edges are shown in  Fig.~\ref{fig:Aq31}. 
Like in the \mbox{6(2,1)-cGNR} case, one identifies a low-energy band of dispersing edge states, now symmetric around $q=0$. 
The different gap position results from the additional zigzag unit for this chirality and the accompanied band folding.\cite{Jaskolski11}
Within MFT, a sizable single particle gap results from the Hubbard interaction, with maximal spectral weight centered at $q=0$. In the QMC data, a similar behavior is observed, albeit with a smaller gap. 
For the \mbox{6(3,1)-cGNR}, we show the local spectral function $A_i(\omega)$ along the ribbon edge in Fig.~\ref{fig:edge3-1local}. In addition to the emergence of the low energy peak at $\omega_\text{max}\approx0.08 t$, this representation also exhibits the modulation of the peak intensity along the majority sublattice sites. The low-energy peak is more prominently observed at the central sites of the zigzag regions and reduces in intensity approaching the armchair regions.
Furthermore, the  MFT  data for both widths in Fig.~\ref{fig:Aq31} display  minima at non-zero $q$ in the low-energy band of $A_\text{edge}(q,\omega)$,
which leads to a bending of the low-energy dispersion near $q=0$.  We observe such a shift in the position of the minimum gap away from $q=0$ and the related bending in the QMC data only for the \mbox{12(3,1)-cGNR}. The gap minimum in the \mbox{6(3,1)-cGNR} still remains at $q=0$, at least within the available resolution. We will comment on this observation further below. 

\begin{figure}[t!]
\centering
  \includegraphics[width=\columnwidth]{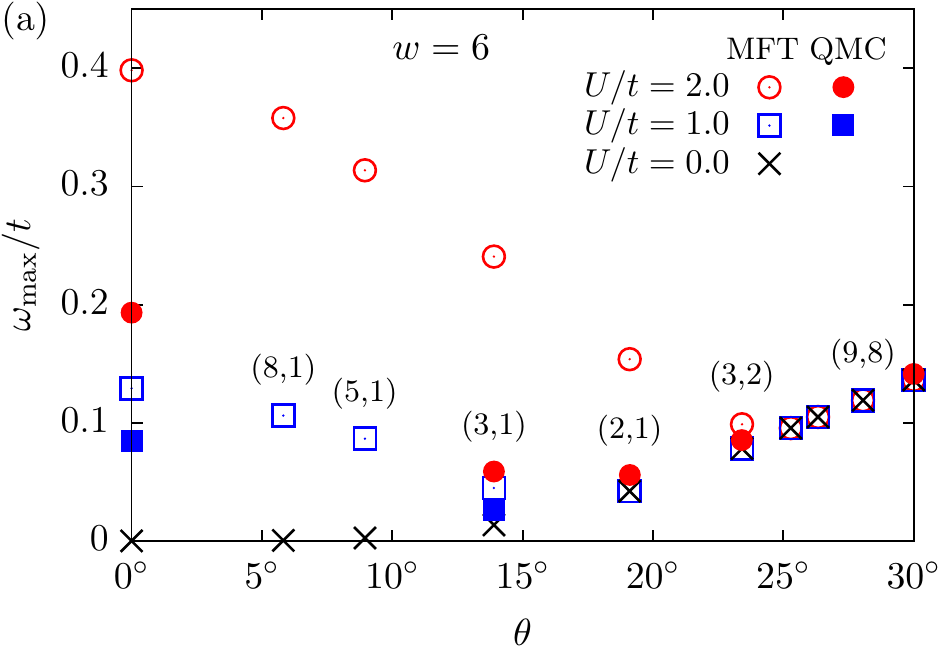}
  \caption{(Color online) Position of the low-energy peak in the edge site LDOS of chiral graphene nanoribbons of width $w=6$ as a function of the chirality angle $\theta$ within MFT (open symbols) and from QMC simulations (filled symbols), both for $U=t$ and $U=2t$. The crosses indicate the single particle gap in the tight-binding ($U=0$) limit.   
  \label{fig:summary6}}
\end{figure}

The results of the LDOS for the different $w=6$ GNRs are summarized in Fig.~\ref{fig:summary6}, which shows the position of the low-energy peak position $\omega_\text{max}$ in $A_\text{edge}(\omega)$ for GNRs of different chiralities. Beyond the QMC and MFT results for the ribbons discussed  above, we included in Fig.~\ref{fig:summary6} MFT results for ribbons of various other chiralities. In addition to the results at $U/t=2$, we also show results for $U/t=1$, as well as the single particle gap of the GNRs in the tight-binding limit, $U=0$. An overall trend towards a reduction of $\omega_\text{max}$ with increasing $\theta$ is observed both in the MFT and QMC data points. MFT however, systematically overestimates $\omega_\text{max}$ -- an effect that apparently is enhanced for increasing chirality. One furthermore observes a sizable increase of the $U=0$  single particle gap due to the finite width of the $w=6$ GNRs for chirality angles $\theta$ beyond about $10^{\circ}$. For larger $\theta$, this leads to a suppression of the edge magnetism already on the MFT level, as mentioned for the \mbox{(3,2)-cGNR} case in the previous section. For example, the \mbox{$(2,1)$-cGNR} and the \mbox{$(3,2)$-cGNR} exhibit no edge magnetism within MFT at $U/t=1$, while for $U/t=2$ the edge magnetism is stable for GNRs below $\theta\approx 25^\circ$. This suppression leads to the apparent merging of the data for the $U=0$ gap and those for $\omega_\text{max}$ at, e.g., $\theta\approx 25^\circ$ within MFT for $U/t=2$.

\begin{figure}[t!]
\centering
  \includegraphics[width=\columnwidth]{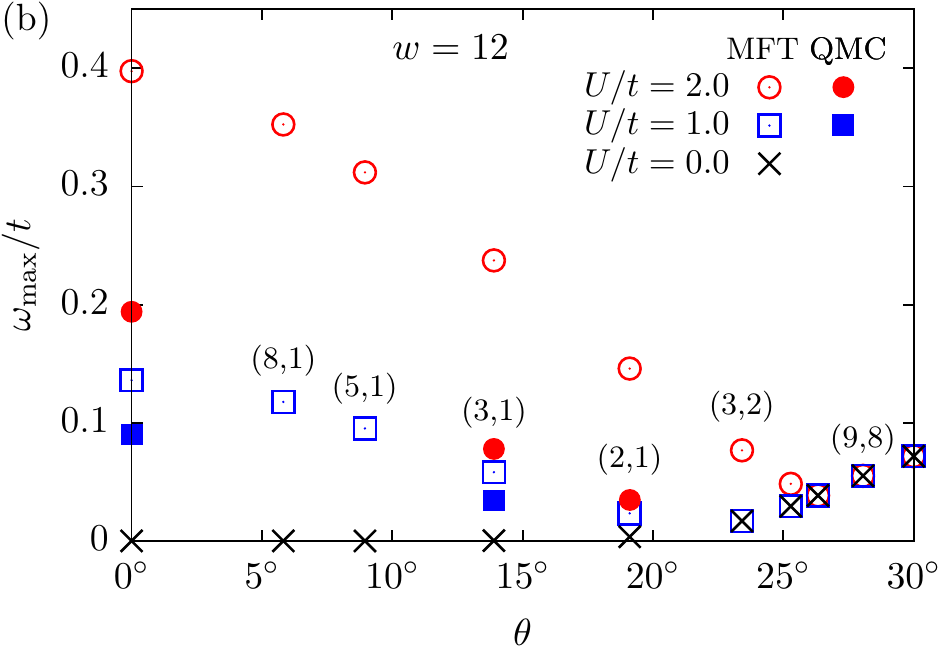}
  \caption{(Color online) Position of the low-energy peak in the edge site LDOS of chiral graphene nanoribbons of width $w=12$ as a function of the chirality angle $\theta$ within MFT (open symbols) and from QMC simulations (filled symbols), both for $U=t$ and $U=2t$. The crosses indicate the single particle gap in the tight-binding ($U=0$) limit.   
  \label{fig:summary12}}
\end{figure}

In Fig.~\ref{fig:summary12}, we show a similar plot for $w=12$ chiral GNRs.
When comparing to Fig.~\ref{fig:summary6}, we find that the position of $\omega_\text{max}$ does not depend strongly on the ribbon width $w$; this holds at least within the chirality range, where the $U=0$ tight-binding gap is small. Such weak $w$-dependence was observed already previously for the zigzag case in Ref.~\onlinecite{Feldner11}, and is also seen for the (3,1)-cGNRs in Fig.~\ref{fig:Aq31}.
The main difference between the two cases, $w=6$ and $w=12$, arises in the large $\theta$-region, where the $U=0$ single particle gap becomes of the order $\omega_\text{max}$, and hence the effective antiferromagnetic inter-edge coupling suppresses the formation of the edge magnetism already within MFT. The intermediate $\theta$-region requires a more careful analysis. 
Consider for example the case of the \mbox{(3,1)-cGNR}: while for $w=6$, the tight-binding gap is already sizable, of the order of the mean-field gap, it is strongly suppressed for the $w=12$ ribbon. This relates to the observation (pointed out above), that for the $w=12$ the gap minimum in $A_\text{edge}(q,\omega)$ shifts away from $q=0$, leading to a bending of the low-energy dispersion near $q=0$. For the \mbox{(2,1)-cGNR}, we do not observe a corresponding bending (at $q=\pi/a$) in the QMC data for  $A_\text{edge}(q,\omega)$ at $w=12$ (not shown), while in MFT such a behavior is already present on the $w=6$ ribbon (cf. Fig.~\ref{fig:Aq31}). In fact, for the \mbox{(2,1)-cGNR} a finite tight-binding gap is still clearly resolved on the scale of  Fig.~\ref{fig:summary12}, and we expect that on even wider ribbons the shift in the gap minimum position and the bending of the dispersion could eventually be detected.

\section{Conclusions}

We examined the interaction-induced edge magnetism in chiral graphene nanoribbons based on a Hubbard-model description using unbiased quantum Monte Carlo simulations. Our results for the equal-time spin-spin correlations confirm that ribbons beyond the armchair limit exhibit substantial ferromagnetic correlations among the zigzag segments along the ribbon edges. Increasing the ribbon width, the effect of the antiferromagnetic inter-edge coupling is strongly suppressed, leading to correlation lengths that are compatible to long-range ferromagnetic edge magnetism. We computed the local spectral functions related to scanning tunneling spectroscopy experiments, and identified a characteristic low-energy peak along the ribbon edge, related to the 
formation of enhanced electronic correlations. The position in energy of this peak is consistent with an essentially linear dependence on the 
interaction 
strength and the chirality angle, and shows no significant width dependence. In future studies, the resilience of these experimentally detectable features  upon the introduction of realistic edge disorder will be investigated. Whether these features prevail as indication for edge magnetism beyond the mean-field approximation, will be most efficiently studied within an effective low-energy spin-only description of the magnetic correlations, which allows for the treatment of significantly larger systems.

\begin{acknowledgments}
We thank F.~F.~Assaad, A.~Honecker, R.~Mazzarello, Z.~Y.~Meng, M.~Morgenstern and M.~J.~Schmidt for discussions. 
Financial support by the DFG under Grant WE 3649/2-1 is gratefully acknowledged, as well as
the allocation of CPU time within  JARA-HPC and from JSC J\"ulich.
\end{acknowledgments}

%
%
\end{document}